\documentclass[twocolumn,floatfix,superscriptaddress,a4paper,showpacs,showkeys,nofootbib,notitlepage]{revtex4-2}
\usepackage[colorlinks=true,linktocpage=true,linkcolor=blue,citecolor=blue,allcolors=blue]{hyperref}
\usepackage{lineno}
\usepackage{epsfig}
\usepackage{latexsym}
\usepackage[utf8]{inputenc}
\usepackage{xspace}
\usepackage{indentfirst}
\usepackage{enumitem}
\usepackage{color}
\usepackage{placeins}

\usepackage{setspace}
\usepackage{lipsum} 
\usepackage{amsmath}
\usepackage{amssymb}
\usepackage[english]{babel}
\usepackage{url}
\graphicspath{{figs/}}
\newcommand{\mean}[1]{\langle #1 \rangle}
\newcommand{\eq}[1]{\begin{align} #1 \end{align}}

\newcommand{\be}{\begin{equation}}
\newcommand{\ee}{\end{equation}}

\begin{document}
\title{Collectivity in $p\text{Pb}$ Collisions with Femtoscopy}
\author{Oleh Savchuk}\thanks{Corresponding author}
\email{savchuk@frib.msu.edu}
\affiliation{Facility for Rare Isotope Beams, Michigan State University, East Lansing, MI 48824 USA}
\date{\today}

\begin{abstract}
Collisions of protons with lead nuclei ($pPb$), such as those measured by the LHCb experiment, provide a unique environment to study the surprising emergence of collective, fluid-like phenomena in small systems. A key signature of this hydrodynamic behavior is the predicted formation of a toroidal vorticity structure. In this work, I use two-particle femtoscopic correlations of non-identical hadrons, specifically proton-pion ($p\pi^+$) pairs,as a novel probe for this phenomenon. Previous works indicate that the collective flow of the system is consistent with the formation of a vortex ring created by the passage of the proton through the lead nucleus, which modifies the collective flow profile. I establish that the resulting emission asymmetry between protons and pions, driven by their mass difference and differential response to the vortical flow, is directly linked to the initial vorticity and can be measured using femtoscopy. This method therefore presents a new, sensitive observable for characterizing the collective dynamics of the matter created in small collision systems.
\end{abstract}


\maketitle

\section{Introduction}\label{Introduction}
The emergence of hydrodynamic behavior in high-energy heavy-ion collisions has become a cornerstone of the field, providing a powerful framework for describing the evolution of the Quark-Gluon Plasma (QGP)~\cite{Jaiswal_2016,shen2020recentdevelopmenthydrodynamicmodeling,GALE_2013}. This paradigm pictures the collision system as an expanding, relativistic fluid where collective flow transforms gradients in the initial geometry and energy density into the anisotropic momentum distributions of final-state particles \cite{hydro-review,Rischke:1995pe,Stoecker:2004qu,Brachmann:1999xt,Brachmann:1999mp,Oliinychenko:2022uvy,PhysRevC.98.014915}. Consequently, significant effort has been devoted to constraining the fundamental properties of the QGP, such as its equation of state (EoS)~\cite{Pratt:2015zsa} and its transport coefficients (e.g., shear and bulk viscosity)~\cite{JETSCAPE:2020mzn}, which regulate relaxation processes toward local thermal equilibrium. Moreover, initial-state fluctuations are now understood as crucial probes of the sub-nucleonic structure of the colliding nuclei~\cite{Savchuk:2024ykb}.

A fascinating challenge to this paradigm arises from the observation that even in "small" collision systems, such as proton-proton ($pp$) and proton-nucleus ($pA$), hydrodynamic models successfully capture the general features of many observables \cite{small-systems,Grosse-Oetringhaus:2024bwr}. This unexpected success has prompted significant theoretical investigation into the mechanisms of rapid thermalization and has spurred the proposal of new, sensitive tests of collectivity. Recently, a unique and inherently hydrodynamic feature—a "smoke ring"—was proposed as a definitive test of collectivity in small systems \cite{smoke-rings}. This phenomenon refers to a toroidal (donut-shaped) vorticity structure predicted to form in central $pA$ collisions, which could be experimentally verified by measuring the polarization of hyperons, such as the $\Lambda$, that couple to this fluid rotation.

While promising, using $\Lambda$ polarization as a direct probe of vorticity faces several challenges. In a medium at local thermal equilibrium, the $\Lambda$ spin is expected to align with the thermal vorticity of the fluid~\cite{Becattini_2020}. However, there is no universal consensus on the most appropriate relativistic definition of vorticity. Furthermore, the QGP is not a perfectly equilibrated system, and deviations from local equilibrium can significantly impact spin-polarization observables. The unknown initial spin distribution within the nuclei introduces an additional, currently unconstrained, source of systematic uncertainty~\cite{Giacalone:2025bgm}. 

In this work, I address these challenges by proposing a complementary approach that utilizes correlations of non-identical particle pairs, specifically protons and pions ($p\pi$)~\cite{zawisza2011meson,Zawisza:2010az}. The recent study at lower collision energies ($\sqrt{s_{NN}} = 2.42~A\mathrm{GeV}$) demonstrated that the distinct emission patterns of pions and protons can be used to link their average spatial displacement to non-radial flow \cite{Savchuk:2025kuk}. In that work, a new observable—the angle between the pair's velocity vector and their relative separation vector—was proposed. The latter is measurable using advanced femtoscopic techniques, often referred to as "source imaging" \cite{femto-imaging}.

The fireball's rotation is not expected to be uniform in low-impact-parameter events; instead, the velocity field should circulate in the predicted toroidal structure. A key signature of this "smoke ring" is that the vorticity vector should change its orientation for particles emitted with opposite momenta relative to the beam axis. Furthermore, a complex transverse dependence of the collective flow is anticipated, as the primary expansion may be deflected by the "lean" of the larger nucleus.

The transverse dependence of longitudinal flow is vital for constraining models of the initial state and understanding the formation of energy and density gradients in the first moments after a collision. Because the lifetime of small systems is short, signatures generated in the initial state are less obscured by viscous dissipation and rescattering \cite{initial-state-small}. Probing the mechanisms of early flow formation is therefore essential for improving the description of heavy-ion collisions across all system sizes. This study outlines a strategy to probe non-radial collective flow and test for toroidal vorticity in small systems using femtoscopic correlations, highlighting the importance of improved 3D initial-state modeling and motivating future experimental efforts.

\section{Femtoscopy}
Femtoscopy aims to experimentally measure the correlation between particle pairs. The correlation function, $C_{\vec{v}}(\vec{q})$, is defined as the ratio of the measured two-particle probability distribution to the product of the single-particle probabilities:
\eq{
C_{\vec{v}}(\vec{q}) =\frac{ \frac{d^6P_{12}}{d^3p_1 d^3p_2} }{\frac{d^3P_1}{d^3p_1}\frac{d^3P_2}{d^3p_2}},
}
where $\vec{p}_i$ is a particle's momentum, $\vec{v}$ is the pair velocity, and $\vec{q}$ is the relative momentum of the particles in the pair's rest frame. This correlation is primarily caused by Final-State Interactions (FSI) and quantum statistics. These effects can be related to the spatial separation of the particles at emission via the Koonin-Pratt formula \cite{Lisa:2005dd}:
\begin{equation}\label{Koonin-Pratt}
C_{\vec{v}}(\vec{q})=\int \mathrm{d}\vec{r}\, K(\vec{q},\vec{r})S_{\vec{v}}(\vec{r}),
\end{equation}
where $S_{\vec{v}}(\vec{r})$ is the source function describing the distribution of relative distances $\vec{r}$ between the particles at emission. The kernel, $K(\vec{q},\vec{r}) = |\psi(\vec{q},\vec{r})|^2$, is the squared relative scattering wave function. A principal goal of femtoscopy is to infer the properties of $S_{\vec{v}}(\vec{r})$ from the experimentally measured $C_{\vec{v}}(\vec{q})$.

The Koonin-Pratt equation relies on several assumptions that are expected to hold reasonably well in high-energy collisions \cite{PhysRevC.56.1095,PhysRevC.111.034903}. To obtain the correlation function, I evaluate the kernel $K(\vec{q}, \vec{r})$ for the $p\pi^+$ pair using the \texttt{Coral} framework \cite{coral}, which incorporates experimentally measured phase shifts to model FSI. For $p\pi^+$, the strong force leads to the $\Delta^{++}$-resonance, while the repulsive Coulomb force produces strong anti-correlation at low relative momenta ($|\vec{q}| \lessapprox 50~\mathrm{MeV/c}$). For non-identical pairs, the source function $S_{\vec{v}}(\vec{r})$ is not generally reflection-symmetric, which translates into an asymmetric correlation function, i.e., $C_{\vec{v}}(\vec{q}) \neq C_{\vec{v}}(-\vec{q})$~\cite{Brown:1997ku}.

\section{Initial Flow Model}
The hydrodynamic approach implies a multistage model of collision dynamics. The evolution begins with a prehydrodynamic stage that generates the initial stress-energy tensor, $T^{\mu\nu}(x)$, and conserved charge distributions. In this work, I use the \texttt{iEBE-MUSIC} hybrid framework with parameters from Ref.~\cite{DobrigkeitChinellato:2024xph}. The initial $T^{\mu\nu}$ is calculated using the \texttt{3dMCGlauber} model, which provides an energy density profile in the transverse plane.

To model the initial longitudinal flow, this 2D profile is boosted according to a rapidity-dependent shift, $y_L(x,y)$. This term models the relative longitudinal motion of "flux tubes" created at different transverse coordinates. It is precisely this relative motion—a longitudinal shear—that generates the initial vorticity field. The flow profile is parameterized as:
\begin{equation}
y_L(x,y) = f \cdot y_{\text{CM}}(x,y),
\label{eq:yL_param}
\end{equation}
where $y_{\text{CM}}(x,y)$ is the center-of-mass rapidity of participant nucleons at a given transverse coordinate $(x,y)$. The parameter $f$ ($0 \le f \le 1$) controls the fraction of this initial rapidity that is converted into the fluid's longitudinal flow. The case $f=0$ recovers the standard boost-invariant picture with vanishing vorticity, while any finite value of $f > 0$ introduces a non-trivial longitudinal shear and, consequently, a non-zero initial vorticity field.

\section{Results}
\begin{figure*}[ht!]
\includegraphics[width=.49\textwidth]{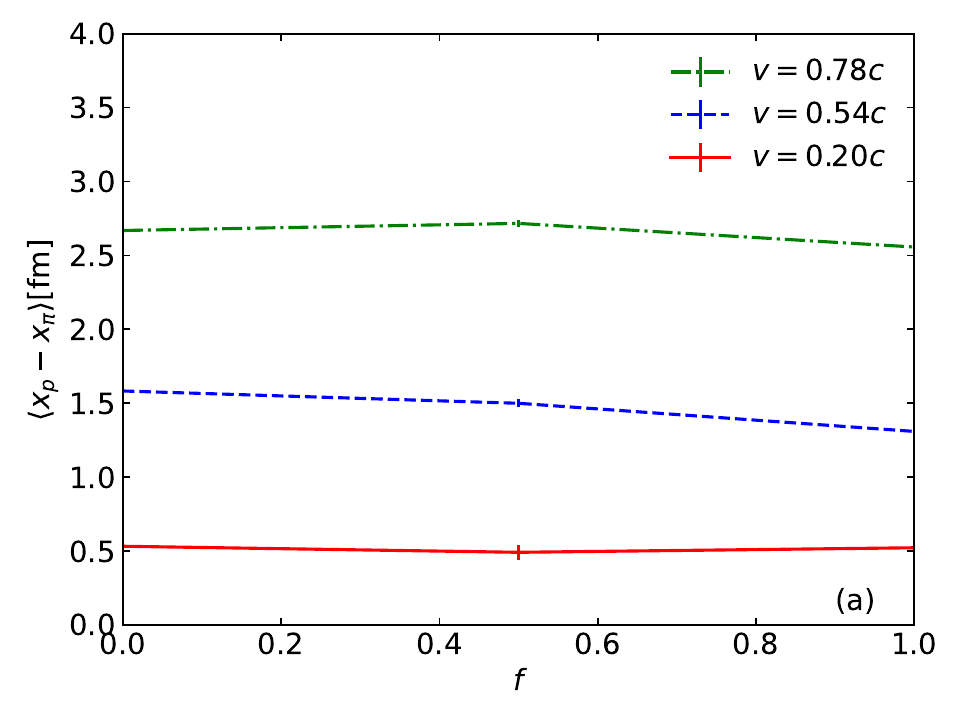}
\includegraphics[width=.49\textwidth]{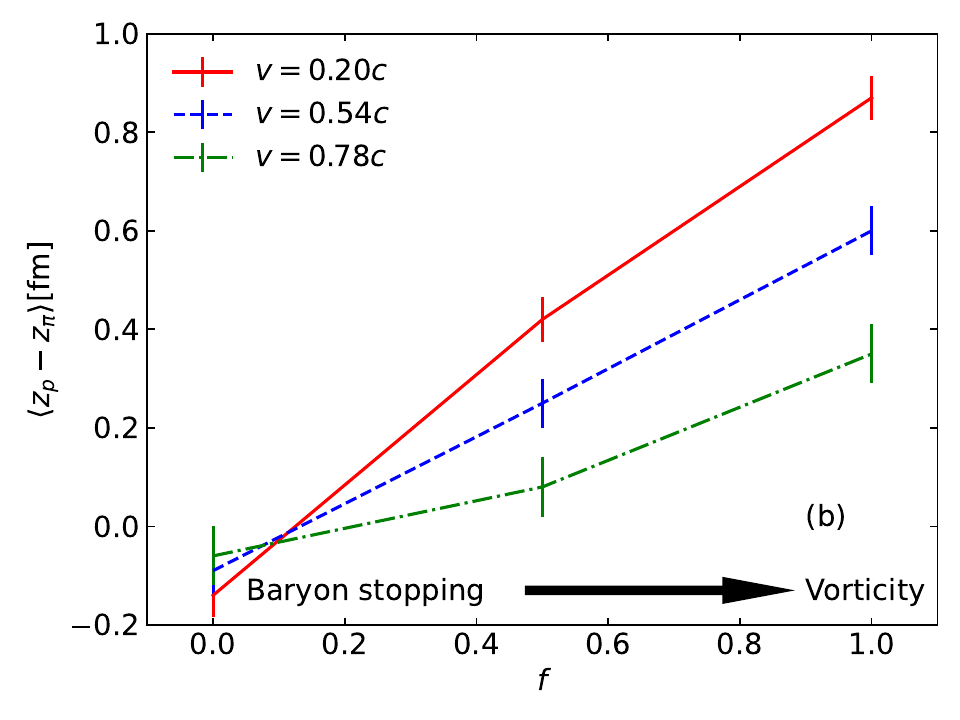}
\caption{\label{fig:displacements}Average relative displacement of proton and pion emission points from simulations with varying initial vorticity (controlled by $f$) for different pair velocities along the $x$-direction. (a) The transverse displacement $\mean{x_p-x_\pi}$ is not sensitive to the longitudinal shear. (b) The longitudinal displacement $\mean{z_p-z_\pi}$ is strongly affected by the initial shear, with a sign change between the $f=0$ (Bjorken-like) and $f=1$ (maximal vorticity) scenarios. This dependence weakens at large pair velocities.}
\end{figure*}

\begin{figure*}[ht!]
\includegraphics[width=.49\textwidth]{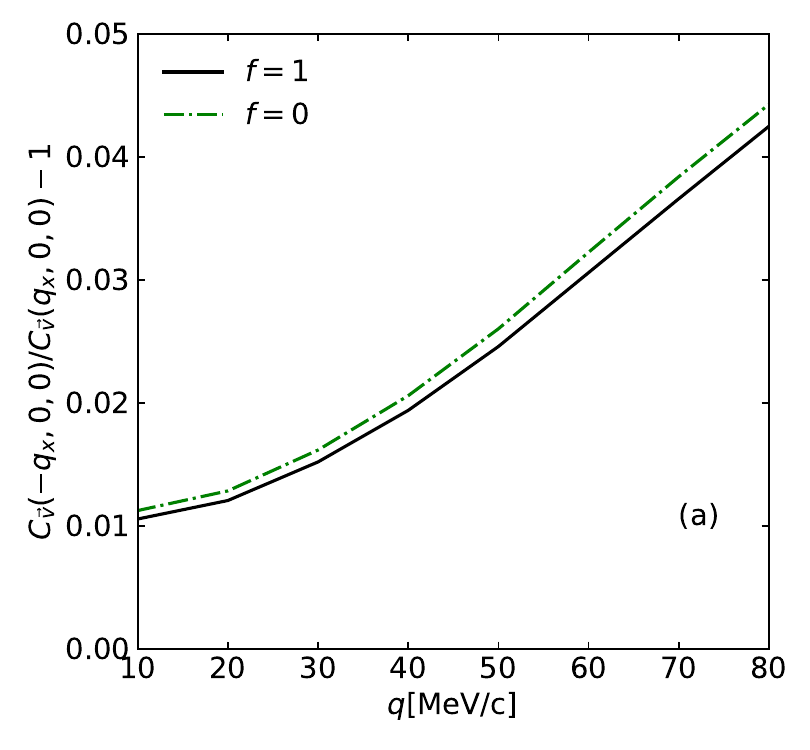}
\includegraphics[width=.49\textwidth]{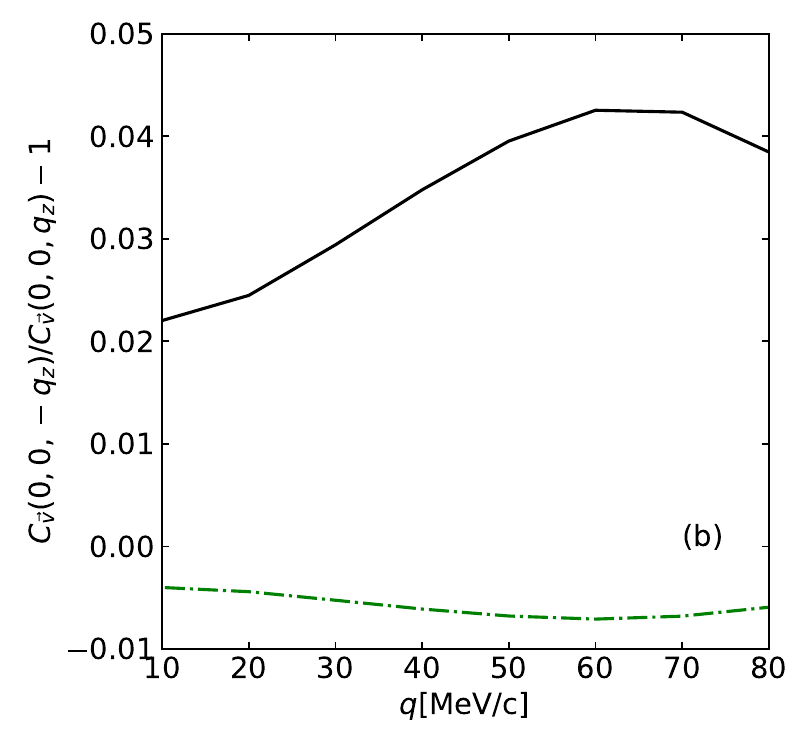}
\caption{\label{fig:Cq-asymmetry}The asymmetry of the correlation function at $v_x=0.20c$, $v_y,v_x=0$, $\frac{C_{\vec{v}}(-q_i)}{ C_{\vec{v}}(q_i)}-1$, caused by the asymmetry of the source. This ratio is directly measurable and can be used to extract the source shifts $\mean{x_p-x_\pi}$ and $\mean{z_p-z_\pi}$ using Eq.~(\ref{eq:asymmetry_ratio}). The sign change in the longitudinal direction (b) between the $f=0$ and $f=1$ cases is a clear signature of vorticity.}
\end{figure*}

\begin{figure}[ht!]
\includegraphics[width=.49\textwidth]{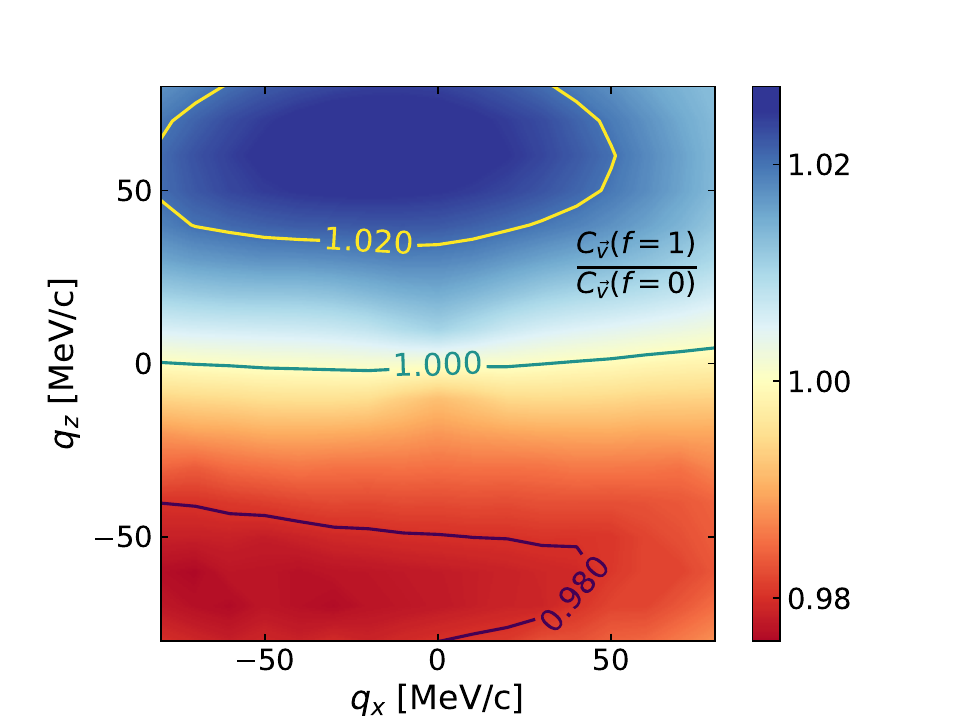}
\caption{\label{fig:Cq-ratio}Ratio of the correlation functions with maximal vorticity ($f=1$) to the case without vorticity ($f=0$) at $q_y=0$. The largest deviations (shown by contours) occur at finite relative momentum ($|q_z| \approx 50~\mathrm{MeV/c}$), making the signal less sensitive to experimental resolution at very low $q$.}
\end{figure}

I analyze $p\pi^+$ correlations in the top 10\% most central $p\text{Pb}$ collisions at $\sqrt{s_{NN}}=72~\mathrm{GeV}$. The key to this measurement is the mass difference between protons and pions. Massive protons ($m_p \approx 938$ MeV) have small thermal velocities compared to their collective velocity and thus act as tracers of the hydrodynamic flow field. In contrast, light pions ($m_\pi \approx 140$ MeV) have large thermal velocities, causing them to be emitted from a broader, thermally smeared region.

The average separation vector, $\langle \vec{r_p}-\vec{r}_\pi \rangle$, between the proton and pion emission points is therefore highly sensitive to the underlying flow profile, which is dictated by the initial-state parameter $f$. As shown in Fig.~\ref{fig:displacements}, This model predicts a clear dependence of the longitudinal separation $\langle z_p-z_{\pi} \rangle$ on $f$.
For the baseline scenario with vanishing vorticity ($f=0$), baryon stopping effects cause protons to lag behind the more centrally produced pions, resulting in a net negative displacement, $\langle  z_p-z_\pi \rangle < 0$.
When the initial shear is maximized ($f=1$), the strong vorticity field imparts a significant forward momentum kick. This effect is mass-dependent, pushing protons more effectively than pions and reversing the displacement to $\langle z_p -z_\pi \rangle > 0$. 
This predicted sign change in the relative longitudinal displacement is a primary finding of This work and serves as a sensitive probe of initial-state vorticity.

A non-vanishing displacement $\langle\vec{r_p}-\vec{r}_\pi\rangle$ corresponds to a source dipole moment, which breaks inversion symmetry and leads to an asymmetric correlation function, $C_{\vec{v}}(\vec{q})\neq C_{\vec{v}}(-\vec{q})$. For a pair interacting via the Coulomb force, this asymmetry can be quantified by the ratio \cite{PhysRevLett.79.4766}:
\begin{equation}
\frac{C_{\vec{v}}(\vec{q})}{C_{\vec{v}}(-\vec{q})}-1 \approx 4\frac{\vec{q}\cdot\langle\vec{r}\rangle}{|\vec{q}|a},\,\, q\to0,
\label{eq:asymmetry_ratio}
\end{equation}
where $a\approx 250~\mathrm{fm}$ is the Bohr radius of the $p\pi^+$ system. This relation provides a direct link between a measurable momentum-space asymmetry and the configuration-space displacement. I apply this formalism to this model results, with the resulting asymmetry $C(-q_i)/C(q_i) -1$ shown in Fig.~\ref{fig:Cq-asymmetry}. Taking values $C(-q_x)/C(q_x) -1\approx 0.01$ at small $q$ one recovers $\mean{x_p-x_{\pi}}\approx 0.6~\mathrm{fm}$. At the same time $C(-q_z)/C(q_z) -1\approx -0.004$ ($f=0$) or $0.02$ ($f=1$) leads to $\mean{z_p-z_{\pi}}\approx -0.25~\mathrm{fm}$ and $1.25~\mathrm{fm}$, respectively. The resulting values appear to be close to the ones obtained from the direct average involving the source function. It could be possible to achieve better reconstruction by considering lower $q<10~\mathrm{MeV}$, however, this will require very good momentum resolution.

The analysis reveals distinct behaviors. The transverse asymmetry is independent of $f$, confirming that $\langle  x_p-x_\pi \rangle$ is insensitive to the longitudinal shear. In stark contrast, the longitudinal asymmetry flips sign between the $f=0$ and $f=1$ cases, directly mirroring the behavior of $\langle z_p-z_\pi \rangle$. To identify the optimal phase space for this measurement, Fig.~\ref{fig:Cq-ratio} shows the ratio of correlation functions at $f=1$ to $f=0$. The strongest effects appear around $q_i \approx 50~\mathrm{MeV/c}$.

The hadronic stage in the $pA$ collision is expected to be rather short with only minor scattering, making it possible to isolate pions produced directly at freeze-out. Heavy resonances are correlated with the flow on the same order as protons and produce pions with a similar degree of correlation. This leads to some reduction of the signal in both $\langle x_p-x_\pi\rangle$ and $\langle z_p-z_\pi\rangle$. This is why in this analysis I ignored resonances with lifetimes greater than 3 fm/c.

The hydrodynamic modeling used in this study lacked initial transverse expansion; therefore, its inclusion could be expected to move the results in velocity. That is, one might expect that the values obtained for the pair velocity of $v=0.2c$ will move to $v\approx 0.5c$. If the resulting flow from proton-nucleus ($pA$) collisions has novel and unexpected features beyond those discussed in this work, the method should still remain valid, allowing for understanding of the flow generated in a collision.

Although there are more involved imaging procedures~\cite{Danielewicz:2006hi, Danielewicz:2005qh,Brown:1997ku}, this simple ratio provides a reliable and computationally efficient estimate of the source's dipole shifts. It is particularly effective in identifying the key physical signature, the sign change of $\langle z_p-z_\pi \rangle$, which serves as a clear indicator of the vorticity of the initial state.

\section{Conclusions}
In this work, I studied proton-pion correlation functions as a sensitive probe of the initial geometry and collective flow in small systems. I argue that systems like $p\text{Pb}$ collisions are ideal for such studies, as their short evolution preserves signatures from the primordial glasma stage with minimal distortion from dissipative effects.

I have shown that femtoscopic correlations of non-identical charged particles can effectively quantify initial fluid vorticity. The spatial asymmetry between the emission points of massive protons (which trace the flow) and light pions (which are thermally diffuse) is translated by final-state interactions into an experimentally measurable asymmetry in their momentum correlation function.

This model identifies two competing effects dictating the pair's longitudinal separation:
1. Baryon stopping (at $f=0$), which pushes protons toward the remnant nucleus, yielding $\langle z_p-z_\pi \rangle < 0$.
2. Vorticity-driven shear flow (at $f=1$), which imparts a strong mass-dependent forward kick, overcoming the stopping effect and flipping the sign to $\langle z_p-z_\pi \rangle > 0$.

This general expectation for the source function of the proton-pion pair is confirmed within the hydrodynamic model. The sign change in the displacement, which is directly accessible through the measure of correlation function asymmetry, makes proton-pion femtoscopy a compelling observable. This methodology is versatile and can be applied to other asymmetric systems, such as electron-nucleus ($eA$) and non-central nucleus-nucleus ($AA$) collisions. Constraining the 3D nature of flow in small systems will lead to more precise initial-state modeling and will complement ongoing efforts to understand spin physics in relativistic fluids.

\begin{acknowledgments}
The author thanks Scott Pratt, Chun Shen for fruitful discussions and contributions. This work was supported by the U.S. Department of Energy, Office of Science, Office of Nuclear Physics under Grant No. DE-FG02-03ER41259.
\end{acknowledgments}

\bibliography{main.bib} 

\end{document}